# Big data in economics

Bogdan OANCEA,
University of Bucharest, Faculty of Business and Administration, Department of Economic and Administrative Sciences,
Bucharest, Romania
bogdan.oancea@faa.unibuc.ro

The term of big data was used since 1990s, but it became very popular around 2012. A recent definition of this term says that big data are information assets characterized by high volume, velocity, variety and veracity that need special analytical methods and software technologies to extract value form them. While big data was used at the beginning mostly in information technology field, now it can be found in every area of activity: in governmental decision-making processes, manufacturing, education, healthcare, economics, engineering, natural sciences, sociology. The rise of Internet, mobile phones, social media networks, different types of sensors or satellites provide enormous quantities of data that can have profound effects on economic research. The data revolution that we are facing transformed the way we measure the human behavior and economic activities. Unemployment, consumer price index, population mobility, financial transactions are only few examples of economic phenomena that can be analyzed using big data sources. In this paper we will start with a taxonomy of big data sources and show how these new data sources can be used in empirical analyses and to build economic indicators very fast and with reduced costs.

1. Introduction

Technological developments in the IT field have led to what we call today the data revolution. The Internet, social networks, smartphones, wearables, different types of sensors are just few of the sources that produce huge volumes of data every day. Online commerce, social interactions, online marketing campaigns, traffic monitoring systems, satellites are many examples of devices and activities that produce data that can be used to describe human and economic behavior.

We now have huge volumes of digital data that are generated by various devices or people's actions and which can be used to describe economic activities, the way people interact or to the policy making process. In order to transform the huge amount of data available today into knowledge of people's or company's behavior, we need to develop new data analysis techniques to deal with unstructured data which require particular attention due to the peculiarities of these data sources.

Processing of these data sets can provide a series of useful information about economic trends, people's behavior or social phenomena. A recent estimate shows that during the next eight years, there will be more than 40 zettabytes of data (Mearian, 2017). This very large quantity of data led researchers to coin a new term - big data. Actually, the term of big data has been used since the 1990s, but it has become very popular around 2012. A recent definition of this term says that big data are characterized by high volume, velocity, variety and veracity that require special analytical methods and software technologies in order to extract value form them (De Mauro et al., 2016). The fifth V has recently been added - Value, that is extracting an added value from them, which is commonly known as big data analytics. For this, it is necessary to precisely identify the new data sources, the best way to collect and process the data to get an added value from them. Working with these new data sources involves a number of issues that need to be considered, such as their quality, integrating with other available data sources, their heterogeneity, their representativeness, the development of special tools for processing large volumes of data, ensuring their confidentiality and security.

There are many papers showing the benefits of using such data sources for companies as well as for governments or society in general (Varian, 2014) but there are only a few proposals for a general framework of using these new data sources to describe and produce forecasts for social and economic activities (Blazquez and Domenech, 2018).

Initially, the term *big data* was used only in the context of parallel and distributed computing technologies such as cloud computing or exascale computing. Today, when we talk about big data, we mean all the data generated in huge quantities in varied formats. There are few works that attempt to build a general framework for collecting, processing, and analyzing large data volumes. As an example, we can mention the work of Pääkkönen, and Pakkala (2015) and Assunção et al. (2015) who speak in this sense of data sources, data collection, data loading and preprocessing, data processing, data analysis, dissemination and visualization of the results. Zhang et al. (2017) proposes another big data architecture to optimize production processes based on four layers: starting with the application of product lifecycle services, then moving to the acquisition of the data from several sources, then processing and storing data and finally using data mining algorithms to extract knowledge from data. Thus, he proposes an architecture with 4 layers: data layer, method layer, result layer and application layer.

Wang et al. (2016) developed a big data architecture in the healthcare field based on five layers: data layer (data sources), data aggregation layer (which is responsible for collecting and storing data), analytics layer (deals with data processing), the information exploration layer (producing outputs for decision support) and finally the data governance layer (managing data throughout for their lifecycle to apply all security and privacy rules).

We have found only few papers proposing a big data framework for the economic and social domain (Blazquez and Domenech, 2018), especially since the particularities of this field imply the existence of a well-defined framework in which these new data sources can be used to produce better decisions or better quality of nowcasting or forecasting of the economic activities.

2. New data sources and analytics

In addition to the traditional data sources used in the analysis of economic activities, such as the official statistics of each country, we are currently witnessing an explosion in the emergence of new data sources. Virtually any human activity is currently leaving digital traces that can be a very useful data source. In the following we present a taxonomy of these data sources (UNECE, 2013):
- Human generated information that represents the record of human experiences: this includes information generated by using social networks like Facebook or Twitter, videos, Internet searches etc.

- Process-mediated data which refer to traditional business systems like financial transactions, e-commerce applications, medical records.
- Machine generated data: these are data generated by different types of sensors (traffic sensors, webcams) mobile/smart phones, satellite images, data generated by intelligent cars, data generated by computer systems, etc.

An important role in the process of generating these new data sets resorts to the Internet. Millions of companies or individuals generate an immense amount of data by using the Internet every day. Search engines on Internet provides services to help researchers to describe, monitor and forecast the economic trends. One example in this area is eloquent: Google Trends has been successfully used to nowcast some macroeconomic variables (Varian, 2104). Forecasts in tourism industry or stock market has also been produced using data from Google Trends (Artola et al., 2015, Preis et al., 2013). Social networks can more or less reflect what is happening at some point in society. Services such as Twitter or Facebook have been used successfully to generate forecasts for economic phenomena or for the study of population mobility (Daas et al. 2012). It should be also taken into account the fact that these data sources have a strong bias towards a particular segment of the population.

Companies' websites provide much information about their business, future intentions, job vacancies, even online services. For example, the analysis of consumer reviews on amazon.com has been used successfully to product demand forecasts. However, the use of information on these websites requires an intense effort since they are unstructured, and it is difficult to standardize the way how the data are collected and processed.

Technological development has led to the unprecedented spread of sensors that are producers of useful data for the analysis of economic activities. Data collected from credit card transactions has been successfully used to predict phenomena such as fraud or bankruptcy. Retail scanners can be used to record everyday purchases and to predict consumer behavior or produce Consumer Price Indices by official statistics bureaus. Data generated by mobile phones has embedded geospatial information and can be very useful to study mobility patterns. Traffic loops can be very useful to predict transportation activities which in turn can be used as a proxy for the overall economic activity.

All of these data sources have some common features: their volume is very large, often they are unstructured or semi-structured, and they are generated at a very high speed. For these reasons, new processing and analysis techniques are needed. Nowadays there are methods based on machine learning techniques that although they have been used in various other fields, we are just beginning to use them in the field of economic analysis.

Since most of the above-mentioned data sources produce semi-structured or unstructured data, in order to be processed they must be brought into a structured form. Much of the effort to incorporate these new sources into data in economic analyzes was directed to the development of data structuring techniques.

If we consider the data collected from the Internet either by web-scrapping techniques or by other techniques, they are in the form of text and, to extract information, they must be processed using special techniques that fall into the field of Natural Language Processing. Some of the techniques specific to this field used to analyze economic activities are Sentiment Analysis, Latent Semantic Analysis or Term Frequency–Inverse Document Frequency.

After data structuring, a data modeling step involving the use of techniques specific to large data sources is required. In this area, machine learning techniques such as supervised learning or unsupervised learning has been widely used (Hastie et al., 2009, Efron and Hastie, 2016). Supervised learning techniques work with data sets that contain both inputs and outputs of an activity and the main purpose is to find a way to infer outputs starting from inputs. These techniques can be further classified into two broad categories: data classification techniques and regression techniques. On the other hand, unsupervised learning methods deal with the situation where we only have inputs and their main purpose is to find relationships between inputs. These kinds of techniques can be divided into clustering techniques and association techniques.

There is no universal recipe for choosing one or another of these techniques, they are usually chosen according to the problem to be solved. For nowcasting or forecasting problems, practical experience has shown that supervised learning methods have delivered the best results. Among these methods, those that have been successfully applied to economic analyses (Varian, 2014) have been decision trees, support vector machines, artificial neural networks, deep learning methods which have proved to be very successful for very large data sets, bagging, boosting, random forests and also the classical linear or logistic regression as well. As an alternative to frequentist statistics approach, the Bayesian statistics approach has been proposed long time ago for economic problems, but the computational complexity of these methods has prevented economists from using them on a large scale. Only in the recent years, as computer performances have increased, and parallel computing has become widespread, the Bayesian statistics based methods become easy to apply. Sometimes, for some models the number of variables is too high and in this case regularization methods are applied (Statistical learning): Least Absolute Shrinkage and Selection Operator or Ridge Regression methods are among the most used techniques in this area (Hastie et al., 2009).

Although a plethora of methods are available to analyze very large datasets, researchers should also consider the robustness of the models used and how accurate the results are. A model is considered robust when it can be generalized to various data sets. Normally, to ensure that the model is robust, the available data set is divided into two subsets: one for training used to compute and adjust the parameters of the model and one for testing which is used to test whether the model achieve a good performance not only for the data used to compute the parameters but for other data sets too. Various techniques exist to fine tune the parameters of the model, such as the cross-validation technique. For the accuracy of the model among the methods used by economists we can mention ROC curves, the Confusion matrices, the Root Mean Square Error (Hastie et al., 2009).

3. Towards a big data methodological framework

To take full advantage of the big data sets available today, some methodological guidelines must be followed. We present in the following some basic elements of a framework that must be implemented to obtain good quality results, sustainable on a long-term basis. Firstly, the access to the data sources should be secured for a long time period. It does not make any sense to invest a lot of resources to obtain some economic indicators for three months for example and then realize that accessing the data is no longer possible. Next, all the phases that data must pass from its collection to visualization of the results must be clearly and transparently defined. There are already some standards proposed in this area, for example the Knowledge Discovery in Databases process (Fayyad et al., 1996), the Cross-Industry Standard Process for Data mining (Chapman et al., 2000) or an adaptation of the Zhang's two phases life cycle model proposed by Salgado et al. (2018).

Analyzing all the proposals regarding the methodological framework that has to be used for big data sources we concluded that at least the following phases should be taken into consideration:

- data collection. This means to ensure all necessary conditions for having access to data sources and use a proper method for collecting the raw data depending on how these data are generated.
- data validation. This phase supposes a complete data checking procedure to answer to the data quality requirements. Some imputation techniques could be also considered during this phase for missing data.
- data documentation. In almost all the cases, big data sources are not meant from the beginning to be used for economic analyses or to measure economic phenomena and they are generated without any metadata. That's why a distinct phase where some documentation is added to the data is certainly needed.
- data transformation: This phase ensures the data are in the proper format for the analysis procedures and if not, it makes the required transformations. This phase could also imply a combination of the big data sources with other available data sets in order to increase the potential for extracting useful knowledge from them.
- data processing/analysis. In this phase all the techniques mentioned in the preceding section could be used, depending on the specific problem to be solved in order to extract knowledge from the data. Specialized software tools are needed in this stage to process very large data sets. Software technologies such as Hadoop, Spark, TensorFlow, NoSql databases etc. are among new trends in processing big data sources.
- visualization. This is the last phase and it implies using software tools to disseminate the results in a way that is understandable not only for researchers but also for normal users. Results should be shared in the form of tables, static charts, interactive graphics or other visual elements that help the users of these results to understand the economic phenomena behind the data or the trends of these phenomena.

4. Conclusions

Nowadays we face a data revolution. Data are produced in enormous quantities by all human activities: from online shopping to the usage of social networks to share information with friends, all human activities generate digital data. These new data sources could be an important opportunity to analyze and understand economic and social trends if they are properly used. In this review article we showed which are the main issues to be tackled with when we what to use unconventional data sources to measure the economic or social phenomena. Starting with data collection and going through metadata generation, data transformation, data analysis and dissemination of the results, we discussed which are the main issues to be considered in order to obtain robust results. Besides the aspects discussed in the previous section, one must consider that most of the new data sources presented here belongs to private companies and accessing the data could interfere with the main business of the data owner. That's why partnerships between the data owners and analysts should be developed in order to have access to the data. The legal and ethical issues should not be neglected too, because most of the data sources provides data at individual level which should be used only under clear and transparent rules.